\documentclass{eas}
\usepackage{graphicx}

\def\aj{AJ}%
\def\araa{ARA\&A}%
\def\apjl{ApJ}%
\def\aap{A\&A}%
\def\mnras{MNRAS}%
\def\pasp{PASP}
\begin{document}

\title{Olivier Chesneau's work on low mass stars} 
\author{Eric Lagadec}\address{Laboratoire Lagrange/Observatoire de la C\^ote d'Azur}

\begin{abstract}
During his too short career, Olivier Chesneau pioneered the study of
the circumstellar environments of  low mass evolved stars using very
high angular resolution techniques. He applied state of the art high
angular resolution techniques, such as optical interferometry and
adaptive optics imaging, to the the study of a variety of objects,
from AGB stars to Planetary Nebulae, via e.g. Born Again stars,  RCB
stars and Novae. I present here an overview of this work and most
important results by focusing on
the paths he followed and key encounters he made to reach these results.
Olivier liked to work in teams and was very strong at linking people
with complementary expertises to whom he would communicate his enthusiasm and
sharp ideas. His legacy will live on through the many people he inspired.
\end{abstract}
\maketitle
\section{Introduction}
During the late stages of their evolution, low to intermediate mass 
(between $\sim$ 0.8 and 8 M$_{\odot}$)
stars develop a large mass-loss (up to 10$^{-4}$M$_{\odot}$yr$^{-1}$)
that plays a key role in the chemical evolution of galaxies by
enriching the interstellar medium with newly formed elements.
This mass loss often leads to the formation of non spherical
circumstellar envelopes. Olivier
Chesneau  became a world leader in the study of these circumstellar
envelopes  using the most advanced high angular resolution techniques,
in order to understand this departure from
spherical symmetry. His work and enthusiastic personality  inspired
many young  scientist, as it was clearly shown during the conference
dedicated to his memory.
I will review his main contributions  to the field of evolved low mass
star.

\section{Binaries and bipolar nebulae}
Low mass stars
appears more or less
spherical in shape on the main sequence. When they reach the planetary
nebula (PN)
phase, the circumstellar envelope of gas ad dust that was formed during
the Asymptotic Giant Branch (AGB phase), and then ejected, gets ionised
by the central star. A wide variety of shapes is observed during this
PN phase, nebulae being round, elliptical, bipolar or
even multipolar. One of Olivier's main interest was to try and
understand this observed symmetry break.

For decades, meetings devoted to the study of the shaping of planetary
nebulae  where the stage of friendly fights between 
{\it magnetic fields}  and {\it binary} people. 
Defenders of magnetic fields could model the formation of the observed
collimated jets, while binary systems could lead to the formation of
equatorial over densities such as disks and torus, that would favour a
polar ejection of matter.  To settle
the debate, a football game was even organised at the
{\it Asymmetrical Planetary Nebulae IV} in La Palma (2007). Olivier took part to the game, but his most important
contribution to the community remains scientific.

Two key papers had an important impact for his research:
\begin{itemize}
\item Soker (2006) showed that a single star can not supply enough energy and angular momentum to shape those nebulae.
\item Nordhaus et al. (2006) demonstrated that magnetic fields can play an important role but isolated stars can not sustain a magnetic field for long enough.
\end{itemize}

Binaries had to play a key role in the shaping of PNe, and Olivier
really wanted to directly detect binaries using high angular
resolution techniques, such as interferometry or adaptive optics
imaging.
Unfortunately,  a given interferometric observation  with two telescopes only probes a given
spatial scale and orientation. To probe different spatial scales, one
needs many baselines (i.e. different projected lengths and angles on the sky
of the line formed by the two telescopes) to fill the so-called
UV plane. This makes it very difficult to detect binaries with
interferometry. For adaptive optics, the difficulties are due to the
limited angular resolution at the time ($\sim$50mas with NACO/VLT in
the K band). Olivier thus realised it would be very difficult to
directly detect binary companions with these techniques (at least with
state of the art instruments around 2005) and decided to look for
indirect
evidences for binaries.

\section{Discs and bipolar nebulae}

Most of the scenarios proposed to explainthe shaping of PNe rely on the presence of
an equatorial disk (see e.g. the review by Balick \& Frank (2002)).
Disks can indeed either channel the mass loss toward the poles 
or lead to the formation of bipolar jets. The presence of such disks
in the core of planetary nebulae can be indirectly observed 
by the presence of dark equatorial lanes in optical observations or
via their polarimetric signatures. However, to better understand the impact of disks on the shaping of the
nebulae, one needs to measure  its mass, size, geometry and angular
momentum. This  can also give us
information about the formation process of the disks. Binary
companion can indeed provide the extra angular momentum necessary for
the stability of a disk.
Olivier decided to combine high angular resolution observations and
radiative transfer modeling to measure the mass and kinematics of such disks.

\section{MIDI/VLTI to peer in the core of Planetary Nebulae}
The disks in the core of planetary nebulae are compact (inner radius
$\sim$ a few AU at 1kpc or more) and dusty.
The best way to directly  peer into their centre and  measure their
size and geometry is achieved by using the highest angular resolution
techniques. Probing their dust emission is also best done in the
mid-infrared regime. Olivier thus used his favourite technique,
namely optical/infrared interferometry to hunt for disks in the core
of  planetary nebulae. The best tool available when he started doing
it was MIDI, the mid-infrared recombiner of the Very Large Telescope
Interferometer (VLTI). Olivier quickly turned MIDI, a two telescope recombiner, into a disk hunting
machine.

\begin{figure}
\centering
\includegraphics[width=10cm]{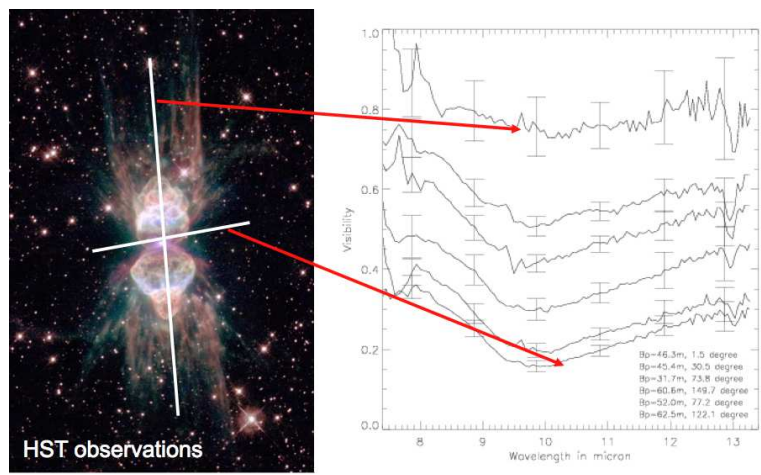}
\caption{Sketch of Olivier's technique to look for disks in the core
  of PNe using interferometry with 2 telescopes. Left: two
  perpendicular baseslines (in whute) as projected on the HST image of
  the PN Menz3el 3. Right: visibilities obtained for different
  baselines orientations. The main idea is to obtain interferometric measurements
  along different  projection angles of the interferometer baseline
  (the line between the two telescopes) on the sky. An interferometer
  measures visibilities, which is the Fourier tranform of the 2D flux integrated perpendicular to the baseline. To translate this
  to people who do not live in the Fourier space (like Olivier) , it
  means that, for a given baseline, if the visibility is large, then
  the measured object is small, and vice versa.  The MIDI/VLTI
  observations of the planetary nebula Menzel 3 presented here show a
  clear interforemtric detection of a disc, as teh visibilities reveal
the presence of a structure flattened in the polar direction.}
\label{inter}
\end{figure}

Optical/infrared interferometry can sound scary to non
interferometrist, but Olivier quickly managed to convince the PN
community that it was a great disc hunting tool.
His observing techniques is quite easy to describe, but more complicated
to apply.  The MIDI instruments measures visibilities along the
baselines formed by the two telescopes it is using. To make it simple,
the visibilities provides an
estimate of the size of the emitting region the interferometer is
probing. Thus, for a given baseline, if a visibility is small, the
object is large and vice-versa. Using different baseline's orientation
(this can be achieved using different telescopes and the earth's
rotation), on can thus measure the size of emitting regions in the
core of PNe along different directions. Fig.\,\ref{inter} shows the interferometric signature of a disk obtained using this technique with MIDI.

\section{Olivier the disk hunter}
\begin{figure}
\centering
\frame{\includegraphics[width=5cm]{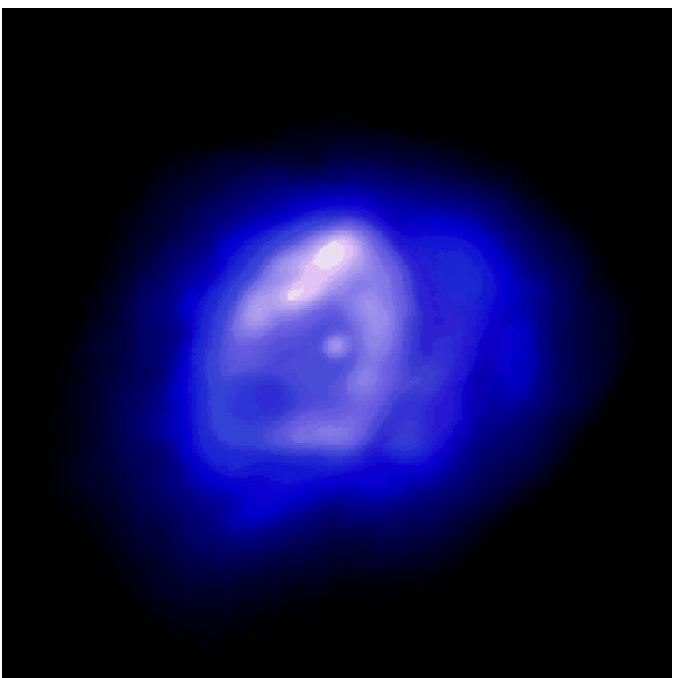}}
\frame{\includegraphics[width=5cm]{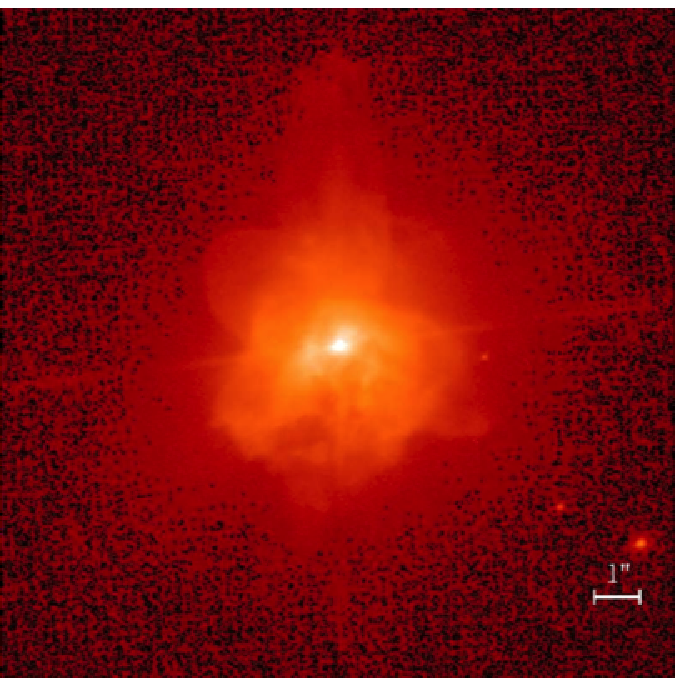}}
\caption{Right: NACO/VLT image of the PN Hen 2-113, revealing the
  presence of an equatorial torus (Lagadec et al., 2006). Right: HST
  image of the PN CPD -568032, displaying a complex morphology with
  jet-like structures in various direction. OLivier resolved a disk in
its core using VLTI/MIDI (Chesneau et al., 2006)}
\label{CPDHen}
\end{figure}

Applying his observing technique with MIDI/VLTI, Olivier became a disk
hunter. A key point of this work was to  combine interferometric observations (to peer
into the core of PNe) with high angular resolution imaging (to study the
larger scale distribution of gas and dust). Having an idea of the
larger scale distribution of gas and dust is indeed very important to
interpret the interferometric signals. His first target was CPD\,-568302, a PN with a [WR] central
star, for which the presence of an edge-on disk was hinted via  HST
spectroscopic 
observations showing an equatorial obscuration (De Marco et al.,
2002). 
He decided to observe this promising target, together with Hen 2-113, 
its {\it sibling} PN, sharing many spectral and spectroscopic
properties. As he was new in the PN field, he contacted Orsola De
Marco who was an expert in the field and had indirectly proven the
presence of a disk in  the core of CPD -568032. One of Olivier's
main quality was to build and lead teams of scientists from differemt
horizons.
For this project he built a strong an enthusiastic team comprising one
of his former PhD supervisor Agnes Acker, Bruce Balick and Albert
Zijlstra. He could find collaborators at any time and in any
situation. I thus joined this project, which ended up being one of the
key point of my career, after meeting Olivier in the restaurant of
the Observatoire de la C\^ote d'Azur in Nice (which is arguably one of
the best restaurant in an astronomical institute worldwide). I was a
PhD student, and Olivier had just been hired as a tenure
astronomer. We quickly realised we both had ESO data on Hen 2-113 and 
decided to work together. I then discovered the {\it Chesneau Method},
defined by three words: enthusiasm, curiosity and passion.
After a month of long days and short (very short) nights, our first
paper together was out.
An equatorial torus was resolved in Hen 2-113
(Fig.\ref{CPDHen}, left; Lagadec et
al., 2006)  from the 
adaptive optics images obtained with NACO/VLT.
A disk was resolved in the core of CPD\,-568032 with the MIDI/VLTI
observations. Interpreting those observations was a real headache
for Olivier and his student Arnaud Collioud, as the structure of CPD\,-568032 (Fig.\ref{CPDHen}) is very
complex. The nebula is made of many jets with different orientations,
and obtaining the geometry of the disk was a real challenge that
Olivier overcame (Chesneau et al., 2006). He combined these
observations with 3D radiative transfer modeling to estimate the
mass of the disk and its dimensions.

Olivier's main lesson from these first successful observations was that
it would be simpler to observe edge-on PNe with clearly defined
bipolar structure. It would then be more straightforward  to determine the
orientation of the central disks. He got inspired by observation by
Smith \& Gehrz (2005) to observe M2-9 (the Butterfly nebula) and Menzel
3 (the Ant Nebula), a work that he did together with his student
Clarie Lykou.
From these VLTI observations, he discovered a disk in the core of
Menzel 3 (Chesneau et al., 2007) and M2-9 (Lykou et al., 2011).
The disk in the core of M2-9 appeared denser and more extended than
the others, so that it was observable with the IRAM interferometer at
plateau de Bure in the millimeter domain. Olivier stared working with
Arancha Castro-Carrizo and showed there was a great synergy between 
infrared and millimeter interferometry observations (Castro-Carrizo
et al., 2012).

\begin{figure}
\centering
\frame{\includegraphics[width=5cm]{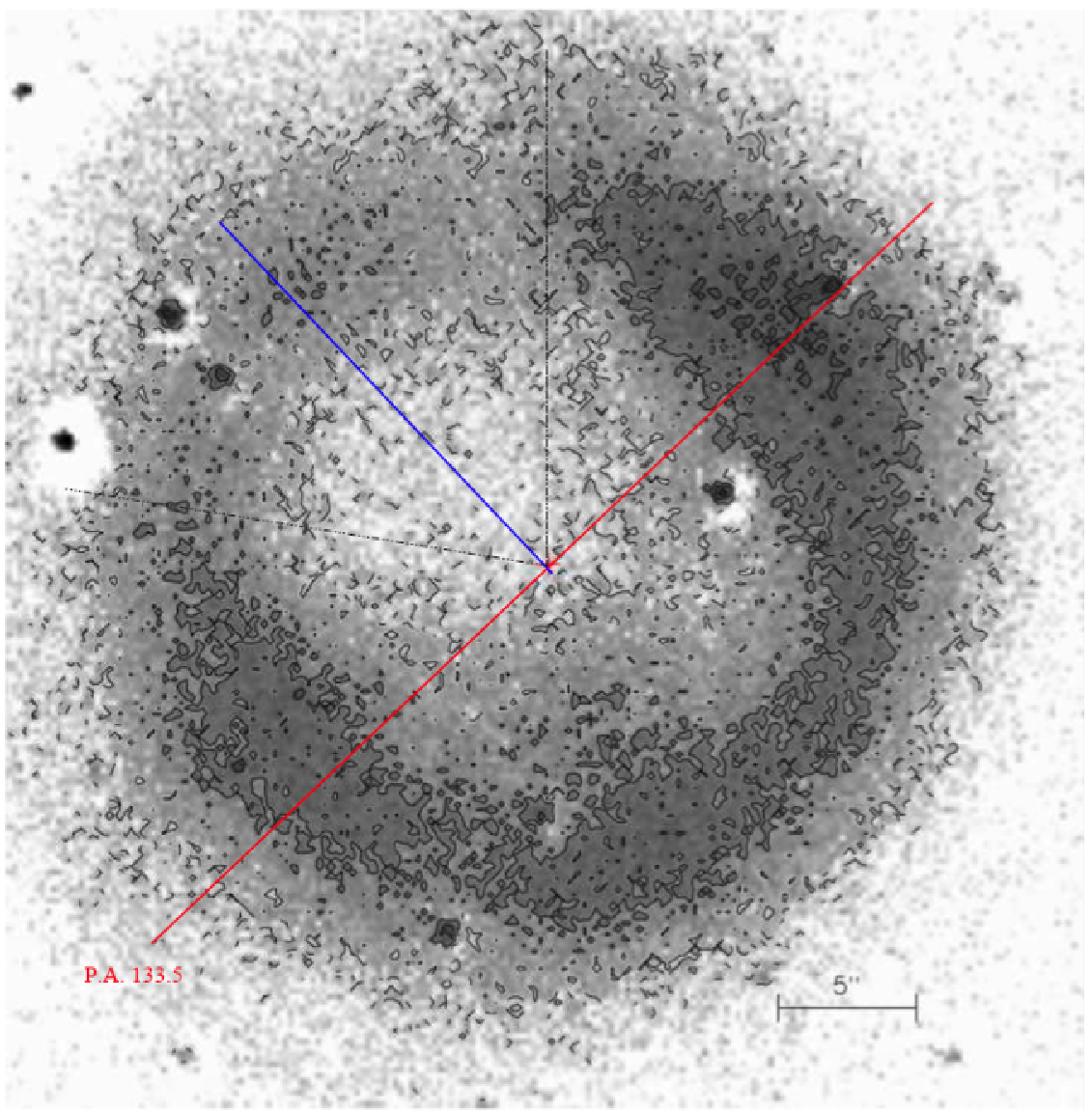}}
\frame{\includegraphics[width=5cm]{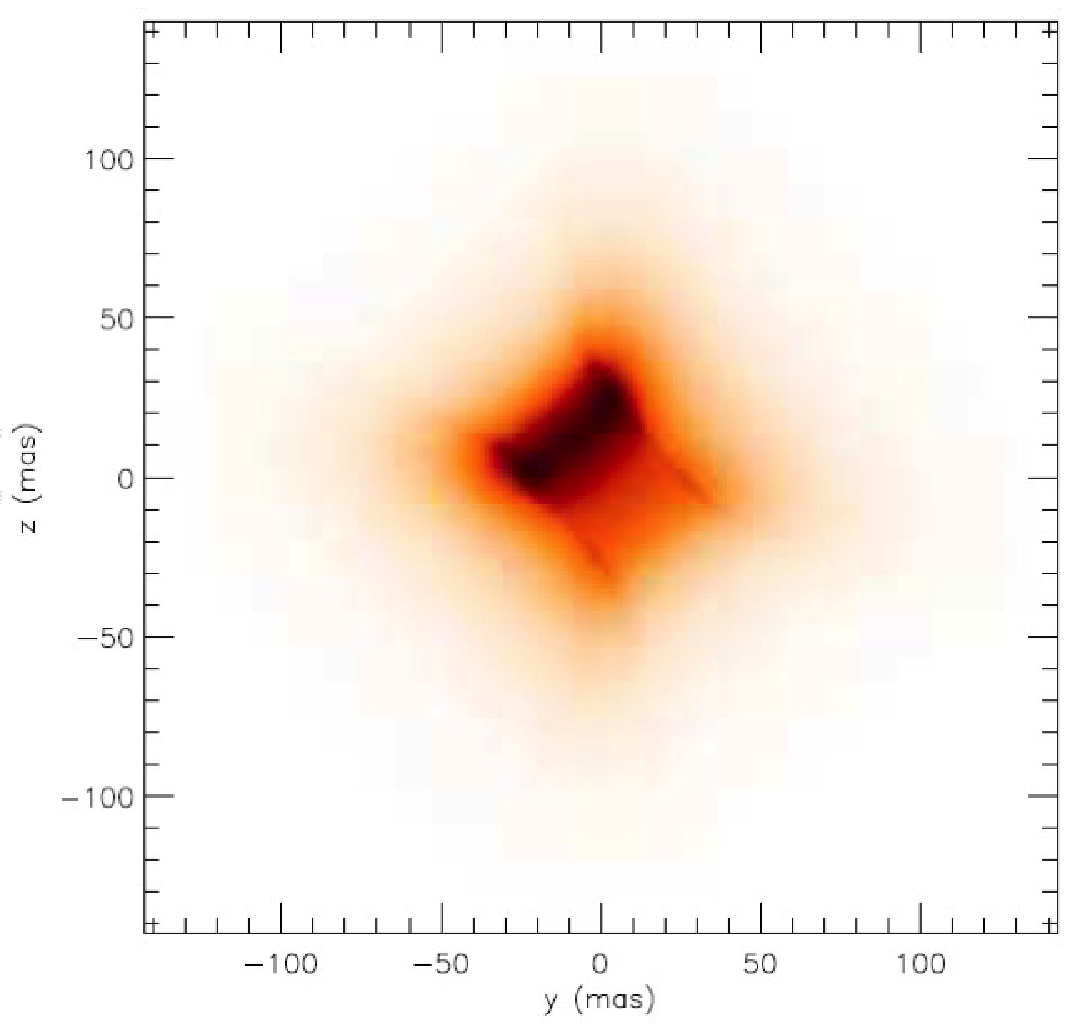}}
\caption{Left: FORS1 VLT [OIII] image of the old PN around the {\it Born
  Again} star Sakurai's object. Right: radiative transfer model image
of the disc found in the core of this object with
MIDI/VLTI. Note that the major axis of the disc is aligned with an
asymetry seen in the old PN (Chesneau et al., 2009)}
\label{sak}
\end{figure}

As Olivier's technique to look for disks in the core of PNe turned out
to be very successful, he
decided
to apply it to new type of objects. Together with Geoff Clayton, he
decided to peer into the core of Sakurai's object, a {\it Born-Again} Star,
i.e. a PN in which the central star underwent a so called {\it Very Late
Thermal Pulse} that reignited nuclear reaction. His observations led to
the detection of a dense equatorial structure (Fig.\,\ref{sak}, left; Chesneau et al., 2009),
that he could not fit with a stratified disk model, as in the case of
PNe.
This is certainly due to the fact that this disk is really young
($\sim$ 15 years). An interesting point of these observations is
that the main axis of the disk is orientated perpendicularly to an
asymmetry observed in the old PN (Fig.\,\ref{sak}, right).
This indicates that the disk/torus  shaping mechanism was also in
action when the AGB star's envelope got ejected.
A few scenarios are possible to explain it, but Olivier was convinced it was due to
the influence of a close binary companion.

His final work was a study of a RCB star, V854 Cen. Those stars are
F-G type supergiants experiencing unpredictable episodes of deep
obscuration  (up to 8 magnitudes in V).  Two scenarios have been
proposed for the formation of RCB stars. They could be the produce 
of the merger of two white dwarfs or due to final helium flash (a bit like
the Born-Again stars) in a PN central star. Olivier wanted to look for
a disk in the core of V854, which would favour the double degenerate
scenario,
as a disk is very likely to be formed around a binary system.
He thus applied his disk hunting technique to V854 Cen. With
observations obtained in 2013, he resolved a structure in
the core of this RCB star, but the  measurements were all made with
baselines with similar orientation. He needed visibilities
measurements with other orientations to determine whether
this structure was spherical or not. While in hospital, he prepared
these observations and we discussed about observations I had obtained
with the mid-infrared imager VISIR/VLT (Lagadec et al., 2011). The
VISIR images were showing the presence of a weak, non spherical,
circumstellar structure. 
However, it was difficult to convince him that the structure seen in
the VISIR images was real and not due to a bad background subtraction.
Together with Djamel M\'ekarnia, we worked on a careful re-reduction
of  the data, until we convincingly showed the presence of an
elongated dusty structure around V854 Cen. Olivier became very
enthusiastic, as the structures seen in the VISIR data were similar to
the one hinted in his MIDI/VLTI data. He just needed data taken 
with new baselines to confirm it. He had a VLTI visitor run the week
after, but he could not go to Chile, as he was already in palliative
care. Florentin Millour prepared the observations with an
over-enthusiastic
Olivier. We recieved a great support from ESO staff in Chile (and
particularly from Willem-Jan de Witt) who helped us obtaining data
without
being able to travel to Chile. The data, reduced within a few hours by
Alain Spang, confirmed Olivier's
prediction,
who then certainly broke a world record by submitting  a paper based on
VLTI data obtained less than 48h before (Chesneau et al., 2014a).

The same week, Olivier worked very actively with Florentin Millour on
data he had just obtained of V838 Mon, the iconic nova-like object
taht erupted in 2002. The eruption was very likely due to a merger
event. 
Olivier's VLTI observations showed that this merger event lead to the
formation of a disk
He submitted this paper  (Chesneau et al., 2014b)  a few hours before
leaving  us to get closer to his stars.

\section{Conclusion}
Olivier Chesneau played a key role for the understanding of the
shaping of bipolar nebulae around low mass stars
via his study of disks in the core of these objects.
His dream was to directly image the binaries that are certainly
responsible for the formation of disks and the  shaping of these objects. To achieve this, Olivier
was very involved in the science team of the  SPHERE/VLT project.
This new generation instrument of the VLT uses an extreme adaptive
optics system to obtain diffraction limited images in the optical with
a resolution down to $\sim$15 mas. Olivier was leading the {\it Other
  Science} group for the SPHERE GTO {\it Guaranteed Time Observations}
team and his main interest in this
project was to map the innermost parts of evolved stars to study their
morphologies, dust formation and hopefully detect binary companions.
Since the first semester of 2015, this instrument is now offered at
the VLT, and one of the first results it delivered was the detection
of a binary companion around the binary star L$_2$ Pup (Kervella et
al., 2015 and this volume). Working with Olivier, Pierre Kervella had identified the
presence of a disk around the AGB star using NACO/VLT imaging
(Kervella et al., 2014). The detection of this binary companion of  an
AGB star was presented during this meeting honouring Olivier's memory.
It was quite symbolic to see one of Olivier's dream being presented there.

Summarizing all the contribution of Olivier Chesneau to the field of
low mass evolved stars was not an easy task, but his main contribution
was to bring a huge leap forward in the study of the innermost part
of circumstellar environment around these stars.
Before Olivier's work, people were talking about {\it sub-arcsec} astronomy,
and Olivier brought the {\it milliarcsec} to the community.

Olivier not only brought state-of-the art high angular resolution
observations to the community, but also ideas about how to make the
best
use of these techniques to understand the physics of these objects.
Olivier was also very good at bringing people with complementary 
expertise to work together and tackle important problems.
His passion and dynamism was a key for that and was immensely
communicative. His main legacy will be this passion and enthusiasm
he gave to a whole generation of young astronomers in Nice and all
over the world.


\end{document}